# Split-gate point-contact for channelizing electron transport on MoS$_2$/h-BN hybrid structures

Chithra H. Sharma and Madhu Thalakulam[a]

*School of Physics, Indian Institute of Science Education and Research Thiruvananthapuram, 695016, Kerala, India*

Electrostatically defined nanoscale devices on two-dimensional semiconductor heterostructures are the building blocks of various quantum electrical circuits. Owing to its atomically flat interfaces and the inherent two-dimensional nature, van der Waals heterostructures hold the advantage of large-scale uniformity, flexibility and portability over the conventional bulk semiconductor heterostructures. In this letter we show the operation of a split-gate defined point contact device on a MoS$_2$/h-BN heterostructure, a first step towards realizing electrostatically gated quantum circuits on van der Waals semiconductors. Our devices show signatures of channelized electron flow and a complete shutdown of transport similar to the conventional point contacts defined on bulk semiconductor heterostructures. We explore the role of back-gate and the drain-source voltages on the pinch-off characteristics and, we are able to tune the pinch-off characteristics by varying the back-gate voltage at temperatures ranging from 4K to 300 K.

Electrostatic gating is a versatile technique to engineer the electron flow in 2D systems. Gate defined quantum point contacts (QPC)[1,2] and quantum dots[3] are the basic building blocks of potential devices for quantum information,[4] quantum metrology[5] and charge sensing applications[6–8]. The inherent non-uniformity in the material properties and the associated spatial variations in the operating conditions make large-scale integration of gated devices on conventional semiconductor heterostructures technically challenging. Heterostructures combining van der Waals (vW) semiconductors and insulators,[9–11] with its atomically flat interfaces and the two-dimensional nature across the entire device area, could provide the required large-scale uniformity. In addition, the substrate independence of vW systems could pave way for flexible and transferable quantum circuits.

Electrostatically defined quantum dots on bilayer graphene-hexagonal boron nitride (h-BN) devices[12] and split-gate defined QPCs on Graphene-Al$_2$O$_3$ structures have been demonstrated.[13] In contrast to graphene, molybdenum disulfide (MoS$_2$)[14] offers superior electrostatically tunable devices owing to the presence of a sizable band gap. Dielectric encapsulation have improved the electron mobility and on-off ratio in MoS$_2$ transistors.[15–18] Compared to conventional oxide interfaces MoS$_2$/h-BN devices offer better electrical properties and stability owing to cleaner interfeces.[19] Quantum transport phenomena such as quantum Hall effect and Shubnikov-de Haas oscillations are observed in MoS$_2$/h-BN devices[20]. In addition, recently, single electron transport has been reported on MoS$_2$[21] making it a potential candidate for hosting future quantum circuits. Dual gating on h-BN/MoS$_2$ has offered better and controllable devices.[9,19] A natural continuation in this regard would be the electrostatic shaping and control of transport on MoS$_2$ based vW heterostructure for nano-electronic applications.

---

[a] Electronic mail: madhu@iisertvm.ac.in

In this communication we demonstrate channelized electron transport on a MoS$_2$/h-BN heterostructure using a split-gate defined point-contact, the first step towards the realization of controllable quantum devices via electrostatic shaping. The MoS$_2$/h-BN heterostructure is formed by aligned transfer of layers of MoS$_2$ and h-BN on a SiO$_2$/Si wafer using a home-built micro-positioning system. Standard electron-beam lithography is used to define the drain and the source contacts and, the split-gate defining the point-contact. We show that the electron flow in the device can be channelized and pinched-off by controlling the voltage on the split-gate defining the point-contact, in the temperature range between 4K and 300 K. We also show that the pinch-off voltage can be continuously tuned by varying the back-gate voltage. Besides these, we also explore the dependence of the pinch-off characteristics of the channel on the drain-source bias.

MoS$_2$ flakes, mechanically exfoliated from bulk crystals are transferred onto a clean 300 nm SiO$_2$/Si substrate and, the drain and the source contacts are defined by electron-beam lithography followed by Cr/Au metallization. A thin flake of h-BN is carefully placed on top of the MoS$_2$ flake using the micro-positioning setup. To ensure a cleaner and a residue-free device both the MoS$_2$ and the h-BN flakes are transferred using the PDMS dry transfer technique.[22] The split-gate defining the point-contact is fabricated on top of the h-BN using electron-beam lithography and Cr/Au metallization. The heavily doped underlying Si substrate is used as the global back-gate for controlling the carrier concentration in the device. All electrical measurements are performed in a 4 K - 300 K variable temperature cryostat in high vacuum (< 10$^{-6}$ mbar) unlit environment.

Fig. 1 (a) shows the optical image of the final device. The top inset shows the optical image of the five-layer MoS$_2$ flake (3.5 nm in thickness) on which the device is made. The Raman spectra shown in the bottom inset exhibit the characteristic $E^1_{2g}$ and $A_{1g}$ peaks of MoS$_2$ verifying the structural quality of the sample. A sketch of the device drawn to-scale in Fig. 1 (b) shows the stacking scheme of MoS$_2$, the drain-source contacts, h-BN and the split-gate. We estimate the thickness of the MoS$_2$ and the h-BN flakes from the AFM height profiles shown in the inset. We have used two flakes of h-BN with a total thickness of ~ 11 nm (5nm + 6nm) to ensure complete coverage of the underlying MoS$_2$ flake. Fig. 1 (c) shows the scanning electron microscope (SEM) image of the device showing the drain and the source contacts and the split-gate defining the point-contact. The lithographic dimensions of the point-contact constriction are 280 nm in length and 220 nm in width. From the optical, AFM and SEM images we infer that the h-BN flake on top of the MoS$_2$ is flat and uniform.

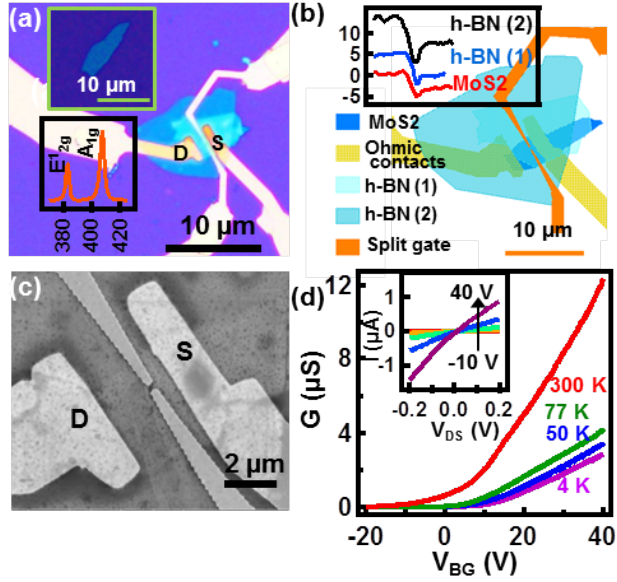

**Figure 1.** (a) Optical image of the device, the drain and the source contacts marked as D and S respectively. The top inset shows the optical image of the MoS$_2$ flake and the bottom inset shows the Raman spectra of the MoS$_2$ showing characteristic Raman peaks. (b) A to-scale cartoon of the device, with all the layers labelled. The inset shows AFM height profiles of the MoS$_2$, h-BN (1) and h-BN (2) flakes showing thickness of 3.5 nm, 5 nm and 6 nm respectively. (c) SEM image of the device showing the point-contact with a lithographic length of 280 nm and width 220 nm. The drain and source contacts are marked as D and S respectively. (d) Conductance versus back-gate voltage at four different temperature. The inset shows I-V characteristics at 300 K, for various back gate voltages.

Fig. 1 (d) shows the conductance of the MoS$_2$ flake versus back-gate voltage ($V_{BG}$) for various temperatures, from 300 K to 4 K with the split-gate grounded. The inset shows I-V characteristics of the device at 300 K for various $V_{BG}$ values. We infer from the I-V characteristics and the conductance versus $V_{BG}$ traces that the sample shows n-type behavior and a linear change in conductance with $V_{BG}$ after the device is turned-on. The threshold voltage estimated from the linear extrapolation for all the temperatures are around $V_{BG}$ = 0 V and, we do not find any significant shift in the threshold voltage as the temperature is lowered, suggesting a lower trap density in our sample.[18]

Now we discuss the formation of the point-contact in our device. For all the measurements discussed in this manuscript we maintain both the electrodes defining the split-gate at the same potential. Fig. 2 (a) shows the pinch-off curve: the current through the device as a function of the voltage on the split-gate ($V_{SG}$), at 4 K (blue), 77 K (green) and 300 K (red). We keep an on-state current (the current through the device while the voltage on the split-gate, $V_{SG}$ = 0V) of 50 nA though the device and $V_{BG}$ of 10 V. The pinch-off characteristics of split-gated point-contacts possess two distinct regimes; (1) the gradual depletion of electrons under the gate resulting in the formation of the point-contact constriction leading to the onset of channelized transport and, (2) the pinch-off of the point-contact itself.[23] We observe a similar behavior in our device as shown in Fig. 2 (a). As $V_{SG}$ is reduced the electrons directly under the gate get depleted (0 V to -3.75 V) resulting in the formation of the constriction exhibiting a shoulder-like structure in the pinch-off characteristics. A further reduction in the $V_{SG}$ results in squeezing the electron flow into the channel followed by the pinch-off of the constriction. This is characterized by the sharp decrease in the conduction followed by a complete shutdown of the transport (-5.75 V and beyond). These features are evident at all the temperatures ranging from 4 K to 300 K. We note here that the features signifying a channelized transport are not observed in a through top-gated device even though the transport could be shut down.[16] The inset to Fig. 2 (a) shows the pinch-off characteristics taken by sweeping $V_{SG}$ in the forward (blue) and reverse (red) directions with an on-state current of 20 nA for $V_{BG}$ = 10 V at 4 K. We do not observe any hysteresis effects in our device suggesting that the charge traps do not play much role in the transport behavior.

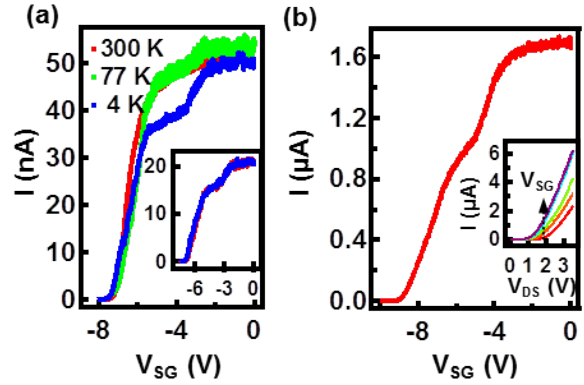

**Figure 2:** (a) Pinch-off curves at 4 K (blue), 77 K (green) and 300 K (red) where the on-state current is kept as 50 nA at $V_{BG}$=10 V. The inset shows pinch-off curves taken in the forward and the reverse direction showing no hysteresis (b) Pinch-off characteristics with an on-state current of 1.7 µA showing a high on-off ratio >$10^6$. The inset shows IV curves as a function of $V_{SG}$ at $V_{BG}$=10 V. $V_{SG}$ is varied from -8 V (red) to 0 V (violet).

Fig. 2 (b) shows the pinch-off characteristics of the point-contact at 4 K with an on-state current of 1.7 µA and $V_{BG}$ = 10 V showing an on-off ratio in excess of $10^6$. We note that by optimizing the $V_{BG}$ and the drain-source voltage $V_{DS}$ (not explored in this work) one can achieve much higher on-off ratios. The inset shows I-V characteristics for $V_{SG}$ starting from 0 V through the pinch-off. Until the point-contact channel formation ($V_{SG} \sim -4$ V), the I-V traces lie one on top of the other suggesting that the entire device area contributes to the transport. As $V_{SG}$ is reduced further through the pinch-off regime, the I-V traces show increasingly flatter off regions suggesting that the transport is dominated by the point-contact barrier.

To get further insight into the pinch-off characteristics we extract the dependence of the pinch-off voltage (the $V_{SG}$ at which the transport shuts down) on $V_{DS}$ and $V_{BG}$. Fig. 3 (a) shows a plot of the pinch-off voltage as a function of $V_{DS}$ at 4 K for $V_{BG}$ of 10 V (green) and 2 V (blue). We find that the pinch-off voltage depends linearly on $V_{DS}$ which is a characteristic behavior of transport across a barrier; the current through the barrier is proportional to the voltage across the barrier and the transmission coefficient of the barrier.[24] As $V_{DS}$ is increased one need to decrease $V_{SG}$ to shutdown the transmission by escalating the barrier. The top inset shows pinch-off curves taken for three different $V_{DS}$, 0.61 V (red), 0.72 V (green) and 0.89 V (blue), for on-state currents of 10 nA, 20 nA and 50 nA respectively through the device. The bottom inset shows a magnified view of the pinch-off region showing a reduction in the pinch-off voltage as $V_{DS}$ is increased while $V_{BG}$ is maintained at 10 V.

Fig 3. (b) shows a surface-plot of the pinch-off characteristics as $V_{BG}$ is varied from 0 V through 10 V in steps of 1 V at 4 K. For all the traces, we have kept an on-state current of 50 nA by varying $V_{DS}$. For all the $V_{BG}$ values we are able to pinch-off the channel and shutdown the transport. As $V_{BG}$ is increased we observe that the pinch-off voltage and the

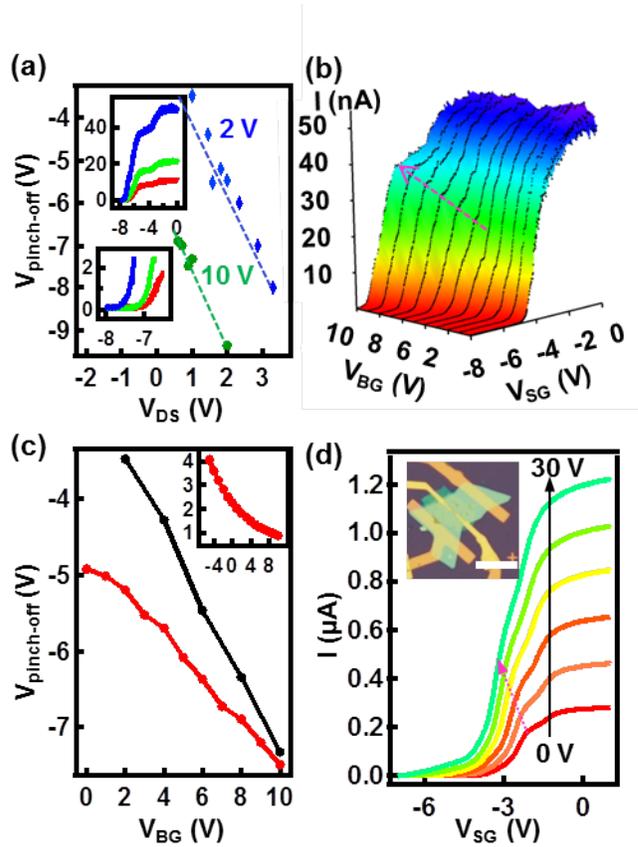

**Figure 3.** (a) Dependence of pinch off voltage on the drain-source voltage for $V_{BG}$ = 2 V (blue) and 10 V (green) at 4 K. Upper inset: pinch-off curves at three different $V_{DS}$, 0.61 V (red), 0.72 V (green) and 0.89 V (blue) with $V_{BG}$ = 10 V. x-axis is $V_{SG}$ in Volts and y-axis is the current through the device in nA. Lower inset: an enlarged view of the pinch-off region. (b) A surface plot of pinch-off curves at 4 K for a series of $V_{BG}$ with the on-state current kept as 50 nA for each trace. The pink arrow shows a linear shift in the position of the shoulder representing the on-set of channelized transport. (c) Variation of the pinch-off voltage with $V_{BG}$. The red trace represents the pinch-off voltage extracted from (b) and the black trace from pinch-off characteristics taken with a $V_{DS}$ of 1 V. The inset shows a plot of the applied $V_{DS}$ for each $V_{BG}$ values to set an on-state current of 50 nA. The x-axis is $V_{BG}$ in volts and y-axis $V_{DS}$ in volts. (d) Pinch-off curves for another point-contact device at 300 K for various $V_{BG}$ values. The arrow indicates the linear shift in the position of the shoulder like structure with the $V_{BG}$. Inset: Optical image of the device, scale bar is 20 μm.

shoulder-like feature moves smoothly towards lower $V_{SG}$ values. For a field effect device, the carrier concentration is proportional to the gate voltage. For point-contacts, the pinch-off voltage varies linearly with the carrier concentration and consequently with the back-gate voltage, which we observe in Fig. 3(c). The black trace represents the pinch-off voltage dependence on $V_{BG}$ while a constant $V_{DS}$ of 1 V is maintained across the device. The red trace represents the pinch-off voltages extracted from the traces in Fig. 3 (b). In this case we have increased $V_{DS}$ as $V_{BG}$ is reduced to keep the on-state current 50 nA for all the pinch-off curves as shown in the inset. We observe that, for the same range of $V_{BG}$ the black trace is steeper than the red trace and the red trace flattens at lower $V_{BG}$ values. Though the reduction in the $V_{BG}$ results in an increase in the pinch-off voltage, the corresponding increment in $V_{DS}$ opposes it, resulting in a reduced slope compared to the black trace and an eventual flattening of the trace.

Apart from this, we observe that the shoulder-like feature in Fig. 3 (b) is more prominent at higher $V_{BG}$. At higher $V_{BG}$, the carrier concentration is higher and the applied $V_{DS}$ is lower, as a result, the depletion of electrons under the gates and the pinch-off of the channel are distinct. Also, the feature moves linearly with $V_{BG}$ as represented by dotted arrows in Fig. 3 (b) as a result of the increase in the carrier concentration.

Fig. 3 (d) shows the room temperature pinch-off curves as a function of $V_{BG}$ for another point-contact device made on a 20 nm thick $MoS_2$ flake. An optical image of the device is shown in the inset. All the pinch off curves exhibit the shoulder-like structure and other features implying a channelized transport similar to the previous device.

In this communication we have demonstrated the formation of split-gated point-contacts and channelized electron flow in a $MoS_2$/h-BN heterostructure, the first step towards the realization of substrate independent quantum circuits to drive the future technology. Our devices exhibit clear and sharp pinch-off behavior and signatures of channelized transport at all temperatures ranging from 4 K to 300 K, unlike the point contacts on conventional semiconductors, making them potential candidates for the implementation of quantum electrical metrology and other charge detection applications at higher temperatures. The devices exhibit transistor action with an on-off ratio in excess of $10^6$. We were able to tune the pinch-off voltage continuously by varying the back-gate voltage. We have studied the effect of carrier concentration and drain-source voltage on the pinch-off characteristics. The devices show nearly Ohmic behavior at room temperature, the drain and the source contacts turned non-linear at lower temperature ranges. Engineering the contacts and the dielectric interfaces would result in better devices exhibiting quantized transport.

**ACKNOWLEDGMENTS**

Authors acknowledge IISER TVM for the infrastructure and the experimental facilities and, Mandar Deshmukh for the help with the micro-positioning setup. MT acknowledges the financial support received from DST-SERB extramural program and CHS acknowledges Abin Varghese for the micromechanical exfoliation of $MoS_2$ in device-2, and CSIR for the fellowship.

**REFERENCES**

[1] D.A. Wharam, T.J. Thornton, R. Newbury, M. Pepper, H. Ahmed, J.E.F. Frost, D.G. Hasko, D.C. Peacock, D.A. Ritchie, and G.A.C. Jones, J. Phys. C Solid State Phys. **21**, L209 (1988).

[2] B.J. van Wees, H. van Houten, C.W.J. Beenakker, J.G. Williamson, L.P. Kouwenhoven, D. van der Marel, and C.T. Foxon, Phys. Rev. Lett. **60**, 848 (1988).

[3] M.A. Kastner, Phys. Today **46**, 24 (1993).

[4] D. Loss and D.P. DiVincenzo, Phys. Rev. A **57**, 12 (1997).

[5] B.J. van Wees, H. van Houten, C.W.J. Beenakker, J.G. Williamson, L.P. Kouwenhoven, D. van der Marel, and C.T. Foxon, Phys. Rev. Lett. **60**, 848 (1988).

[6] J.M. Elzerman, R. Hanson, L.H. Willems van Beveren, B. Witkamp, L.M.K. Vandersypen, and L.P. Kouwenhoven, Nature **430**, 431 (2004).

[7] M. Field, C.G. Smith, M. Pepper, D.A. Ritchie, J.E.F. Frost, G.A.C. Jones, and D.G. Hasko, Phys. Rev. Lett. **70**, 1311 (1993).

[8] M. Thalakulam, C.B. Simmons, B.M. Rosemeyer, D.E. Savage, M.G. Lagally, M. Friesen, S.N. Coppersmith, and M.A. Eriksson, Appl. Phys. Lett. **96**, 183104 (2010).

[9] G.-H. Lee, Y.-J. Yu, X. Cui, N. Petrone, C.-H. Lee, M.S. Choi, D.-Y. Lee, C. Lee, W.J. Yoo, K. Watanabe, T. Taniguchi, C. Nuckolls, P. Kim, and J. Hone, ACS Nano **7**, 7931 (2013).

[10] S. Das, R. Gulotty, A. V. Sumant, and A. Roelofs, Nano Lett. **14**, 2861 (2014).

[11] A.K. Geim and I. V Grigorieva, Nature **499**, 419 (2013).

[12] A.M. Goossens, S.C.M. Driessen, T.A. Baart, K. Watanabe, T. Taniguchi, and L.M.K. Vandersypen, Nano Lett. **12**, 4656 (2012).

[13] S. Nakaharai, J.R. Williams, and C.M. Marcus, Phys. Rev. Lett. **107**, 36602 (2011).

[14] K.F. Mak, C. Lee, J. Hone, J. Shan, and T.F. Heinz, Phys. Rev. Lett. **105**, 136805 (2010).

[15] H. Liu and P.D. Ye, IEEE Electron Device Lett. **33**, 546 (2012).

[16] B. Radisavljevic, A. Radenovic, J. Brivio, V. Giacometti, and A. Kis, Nat. Nanotechnol. **6**, 147 (2011).

[17] B. Radisavljevic and A. Kis, Nat. Mater. **12**, 815 (2013).

[18] W.S. Leong, Y. Li, X. Luo, C.T. Nai, S.Y. Quek, and J.T.L. Thong, Nanoscale **7**, 10823 (2015).

[19] G.-H. Lee, X. Cui, Y.D. Kim, G. Arefe, X. Zhang, C.-H. Lee, F. Ye, K. Watanabe, T. Taniguchi, P. Kim, and J. Hone, ACS Nano **9**, 7019 (2015).

[20] X. Cui, G.-H. Lee, Y.D. Kim, G. Arefe, P.Y. Huang, C.-H. Lee, D.A. Chenet, X. Zhang, L. Wang, F. Ye, F. Pizzocchero, B.S. Jessen, K. Watanabe, T. Taniguchi, D.A. Muller, T. Low, P. Kim, and J. Hone, Nat. Nanotechnol. **10**, 534 (2015).

[21] K. Lee, G. Kulkarni, and Z. Zhong, Nanoscale **8**, 1271 (2016).

[22] A. Castellanos-Gomez, M. Buscema, R. Molenaar, V. Singh, L. Janssen, H.S.J. van der Zant, and G. a Steele, 2D Mater. **1**, 11002 (2014).

[23] H. van Houten, C.W.J. Beenakker, and B.J. van Wees, *Semiconductors and Semimetals* (Elsevier, 1992).

[24] D.K. Ferry, S.M. Goodnick, and J. Bird, *Transport in Nanostructures*, 2nd edn (Cambridge University Press, n.d.).